\documentclass[twocolumn,aps,prb]{revtex4}
\usepackage[T1]{fontenc}
\usepackage[latin1]{inputenc}
\usepackage{amsmath}
\usepackage{amstext}
\usepackage{epsfig}
\usepackage{epic}
\usepackage{eepic}
\usepackage{amssymb}
\usepackage{exscale}
\usepackage{pst-node,array}
\begin{document}
\title{A self-consistent perturbative evaluation of ground state energies: application to cohesive 
energies of spin lattices}
\author{Mohamad Al Hajj and Jean-Paul Malrieu}
\affiliation{Laboratoire de Physique Quantique, IRSAMC/UMR5626, Universit\'e Paul Sabatier, 118 route 
de Narbonne, F-31062 Toulouse Cedex 4, France}
\begin{abstract}
The work presents a simple formalism which proposes an estimate of the ground state energy from a 
single reference function. It is based on a
perturbative expansion but leads to non linear coupled equations. It can be viewed as well as a 
modified coupled cluster formulation.
Applied to a series of spin lattices governed by model Hamiltonians the method leads to simple 
analytic solutions. 
The so-calculated cohesive energies are surprisingly accurate. Two examples illustrate its 
applicability to locate phase transition.
\bigskip
\end{abstract}
\maketitle
\section{Introduction}
The quantum many-body problem has been perfectly clarified a long time ago for the single-reference 
approaches. Its logics is given by 
the linked cluster theorem,\cite{Ref1,Ref2,Ref3} which relies on a perturbative expansion and its 
diagrammatic representation. A panoply of 
computational methods satisfying the basic size-consistency requirements are available. One of them 
is the M\"oller-Plesset order-by-order 
perturbative expansion \cite{Ref4} which suffers from its slow convergence (and in some cases 
high-order divergences \cite{Ref5}).
Other methods lead to non-linear equations coupling the coefficients of the vectors which are 
singly or doubly excited with respect
to the single reference $\Phi_{0}$. Among them one may quote the Coupled Electron Pair Approximation 
(CEPA) formalism \cite{Ref6,Ref7,Ref8} which 
eliminates unlinked diagrams from a Singles and Doubles Configuration Interaction (SDCI) and 
introduces part of (in CEPA-2 \cite{Ref7,Ref8}) or all 
(in the $\mbox{SC}^{2}$ SDCI version \cite{Ref9}) of the Exclusion Principle Violating diagrams.
The Coupled Cluster Method (CCM \cite{Ref10,Ref11,Ref12}) has a better standard since it is an 
elegant and 
generalizable formalism. This method gives an 
exponential structure to the wave operator $\Omega$ which transforms a reference function $\Phi_{0}$ 
in the exact solution $\Psi_0$, 
$\Psi_0=\Omega\Phi_{0}$, $\Omega=exp(S)$. In practice $S$ in limited to a certain set of excitations. 
The simplest version, 
called CCSD, introduces single and double excitations on the top of $\Phi_{0}$ and takes into 
account the connected effect of the 
quadruples. It does notes treat correctly the effect of the Triples, which has to be incorporated 
perturbatively in the CCSD(T) formalism.\cite{Ref13,Ref14}
\\
The present work returns to a low-order semi-perturbative development from $\Phi_{0}$. It leads to a 
system of equations fixing
the coefficients of the singly and doubly excited vectors, which
\begin{itemize}
\item[-] does not require to assume an exponential structure of the wave-operator,
\item[-] introduces the effect of deviations from additivity of excitation energies (deviations which 
is implicitly supposed to be negligible in CCSD),
\item[-] and introduces some high-order effects through EPV corrections to the excitation energies.
\end{itemize}
The method is applied to a few infinite periodic spin lattices governed by Heisenberg Hamiltonians 
which only consider interactions between 
nearest neighbors.   
In these problems if all bonds are identical, the solutions of our self-consistent perturbative (SCP) 
method lead to  simple polynomial equations. The results, compared to the exact solutions, are of 
surprising quality. The method is then employed to identify phase transitions in two different 2-D
problems.
\section{Formalism}
\subsection{Generalities}
Let us call $\Phi_{0}$ a suitable single reference supposed to be a relevant approximation to the
ground state of the system governed by an Hamiltonian $H$ and $\Phi_{m}$ vectors orthogonal to 
$\Phi_{0}$.
In the intermediate normalization
\begin{equation}
\Psi_{0}=\Phi_{0}+\sum_{m}C_{m}\Phi_{m},
\end{equation}
the exact energy is given by the eigenequation relative to $\Phi_{0}$, 
$\langle\Phi_0\vert H-E\vert \Psi_0\rangle=0$
\begin{equation}
E=H_{00}+\sum_{k}H_{0k}C_{k},
\end{equation}
where $H_{ij}=\langle\!\Phi_{i}\vert H \vert\Phi_{j}\!\rangle$. 
The Hamiltonians is at most bi-electronic. One may call $h_1$ and $h_2$ the mono-electronic and 
bi-electronic parts of the Hamiltonian $H=h_1+h_2$. One may label $m,n...$ (holes) the mono-electronic 
function occupied in $\Phi_0$ and $r,s...$ (particles) those which are not occupied in $\Phi_0$.\\
In the following we shall refer to an Epstein-Nesbet\cite{Ref15,Ref16,Ref17} zero-order Hamiltonian 
which is the diagonal
part of the Hamiltonian in the basis of the N-electronic vectors $\{\Phi_0,..\Phi_m..\}$
\begin{equation}
H_0 = \sum_{m}\vert\Phi_{m}\rangle \langle\Phi_{m}\vert H\vert\Phi_{m}\rangle\langle\Phi_{m}\vert.
\end{equation}
The perturbation operator $V$
\begin{equation}
V=H-H_0
\end{equation}
is zero-diagonal in this basis
\begin{equation}
V=\sum_{i}{\sum_{j}}^{\prime}\vert\Phi_{i}\rangle \langle\Phi_{i}\vert H\vert\Phi_{j}\rangle\langle\Phi_{j}\vert.
\end{equation}
The Hamiltonian being at most bi-electronic, it appears from Eq. 2 that the knowledge of the 
coefficient of the singly and doubly excited vectors which interact with $\Phi_0$ is sufficient to 
give the ground state energy. We shall call first generation $S_{1}$ the set of vectors $\Phi_i$
interacting with $\Phi_0$, 
\begin{equation}
S_1=\{\Phi_i\},  \hspace{0.5cm} \langle\Phi_{i}\vert H\vert\Phi_{0}\rangle\not=0 \nonumber
\end{equation}
and $T^{+}_{i}$ the operators creating the vectors $\Phi_i$ from $\Phi_0$, 
\begin{equation}
\Phi_{i}=T^{+}_{i}\Phi_{0}. \nonumber
\end{equation}
These operators may be written as 
\begin{equation}
T^{+}_{i}=a^{+}_{r}a_{a}(1-\delta(\langle r\vert h_1\vert a \rangle))
\end{equation}
or 
\begin{equation}
T^{+}_{k}=a^{+}_{r}a^{+}_{s}a_{b}a_{a}(1-\delta(\langle rs\vert h_2\vert ab \rangle))
\end{equation}
where $\delta$ is the Kroneker symbol.
The coefficients of the first generation vectors can be estimated from the eigenequations relative 
to them. For $\Phi_{i}$ the equation $\langle\Phi_i\vert H-E\vert \Psi_0\rangle=0$ can be written 
using Eq. 2 as 
\begin{eqnarray}
 & &\left(
   H_{ii}-H_{00}-\sum_{k}H_{0k}C_{k}
 \right)
  C_{i}+H_{i0}
\nonumber \\ 
& &+\sum_{j\in S_{1}}H_{ij}C_{j}+\sum_{\alpha \not\in S_{1}}H_{i\alpha}C_{\alpha}=0.
\end{eqnarray}
To determine the coefficients $C_i$ of the first generation vectors, it is sufficient to have
an estimate of the coefficient of vectors $\Phi_{\alpha}$ of the second generation, i.e., 
those which interact with the vectors of $S_1$.
Among the vectors $\Phi_{\alpha}$ belonging to the second generation one may distinguish those 
which are obtained from $\Phi_{i}$ by the actions of 
operators $T^{+}_{k}$ possible on $\Phi_{0}$ (type 1), and the others which are obtained from $\Phi_{i}$ by 
operators $R^{+}_{m}$ different from the 
${T'^{+}_{s}}$ (type 2). \\
If one calls $T$ the sum of the operators $T^{+}_{k}$ and of their adjoints 
\begin{equation}
T=\sum_{k}(T^{+}_{k}+{T^{+}_{k}}^{\bot})
\end{equation}
the operators $R$ 
\begin{equation}
R=1-\sum_{l}\vert \Phi_{l}\rangle \langle\Phi_{l}\vert -T,
\end{equation}
is the sum of all the operators changing one or two mono-electronic functions which are different 
from $T^{+}_{k}$ operators. Notice that $\langle R^{+}_{m} \Phi_{0}\vert H \vert \Phi_{0} \rangle=0$.
With the Hamiltonians considered hereafter the second generation vectors $\Phi_{\alpha}$ are either 
of type 1 or of type 2, i.e., are generated either as 
\begin{equation}
\Phi_{\alpha}=T^{+}_{l}T^{+}_{k}\Phi_{0} \hspace{0.5cm} \mbox{(type 1)} \nonumber
\end{equation}
or as 
\begin{equation}
\Phi_{\alpha}=R^{+}_{m}T^{+}_{k}\Phi_{0} \hspace{0.5cm} \mbox{(type 2)} \nonumber
\end{equation}
as pictured in Fig. 1.
\begin{figure}[h]
\unitlength=1mm
\begin{picture}(80,33)
\linethickness{0.1mm}
\put(50,5){$\Phi_{0}$}
\put(40,15){$\Phi_{n}$}
\put(60,15){$\Phi_{m}$}
\put(50,25){$\Phi_{\alpha}$}
\put(50,7){\line(-1,1){8}}
\put(52,7){\line(1,1){8}}
\put(42,17){\line(1,1){8}}
\put(60,17){\line(-1,1){7.5}}
\put(31,15){$\Phi_{k}$}
\put(68,15){$\Phi_{i}$}
\put(78,15){$\Phi_{l}$}
\put(73,25){$\Phi'_{\alpha}$}
\qbezier(70,17)(71.5,21)(73,25)
\qbezier(77,25)(78.5,21)(80,17)
\qbezier(53.5,6)(61,10.25)(68.5,14.5)
\qbezier(48.5,6)(40.25,10.25)(32,14.5)
\qbezier(54,6)(66,10.5)(78,15)
\put(66,20.5){$R_{m}^{+}$}
\put(79,20.5){$R_{n}^{+}$}
\put(61.5,11){$T_{i}^{+}$}
\put(35.5,8){$T_{k}^{+}$}
\qbezier(53.5,26)(61,21.75)(68.5,17.5)
\qbezier(48.5,26)(40.25,21.75)(32,17.5)
\put(60,23){$T_{k}^{+}$}
\put(36,22){$T_{i}^{+}$}
\put(70,9){$T_{l}^{+}$}
\put(43,10){$T_{n}^{+}$}
\put(52,12){$T_{m}^{+}$}
\put(43,20){$T_{m}^{+}$}
\put(52,18){$T_{n}^{+}$}
\put(1,5){Reference}
\put(1,15){First generation}
\put(1,25){Second generation}
\put(47,30){(type 1)}
\put(70,30){(type 2)}
\end{picture}
\caption{Illustration of the genealogic generation of $\Psi_0$ from the reference function $\Phi_0$}
\end{figure}
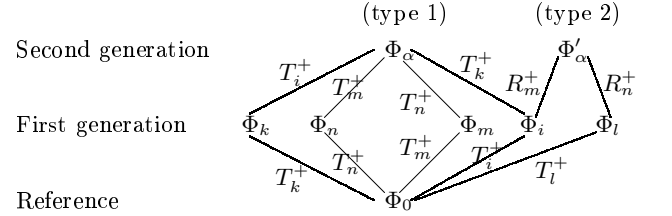
\subsection{Evaluation of the coefficients of second generation vectors}
Regarding the \textit{vectors of type 1}, the operators $T^{+}_{k}$ are in general possible on 
$\Phi_{i}$, but some of them are impossible ($T^{+}_{k}\Phi_{i}=0$). Let call $EPV(i)$ 
(Exclusion Principle Violating correction) the quantity\cite{Ref9} 
\begin{equation}
EPV(i)=\underset{T^{+}_{k}\Phi_{i}=0}{\sum_{k}}H_{0k}C_{k}.
\end{equation}
Noting that if $\Phi_{\alpha}=T^{+}_{k}\Phi_{i}$, and if $T^{+}_{k}$ is a double excitation 
$H_{i\alpha}=H_{0k}$. Introducing Eq. 11 the Eq. 8 can be written 
\begin{eqnarray}
& &\left(
H_{ii}-H_{00}-EPV(i)
\right)  
C_{i}+H_{i0} 
\!+\!\sum_{j \in S_{1}}H_{ij}C_{j} 
\nonumber \\
& & + \underset{T^{+}_{k}\Phi_{i}\not=0}{\sum_{k}H_{0k}}
\left(
  C_{T^{+}_{k}\Phi_{i}}-C_{i}C_{k}
\right) 
\nonumber \\
& & + \underset{R^{+}_{m}\Phi_{i}\not\in S_{1}}{\sum_{m}}\langle\!\Phi_{i}\vert H\vert R^{+}_{m}\!\Phi_{i}\!\rangle C_{R^{+}_{m}\Phi_{i}}\!=\!0.
\end{eqnarray}
A first order evaluation of $C_{i}$ is of course the Epstein-Nesbet\cite{Ref15,Ref16,Ref17} one 
\begin{equation}
C_{i}=H_{i0}/(H_{00}-H_{ii}).
\end{equation}
Introducing $EPV$ corrections in the energy denominator
\begin{equation}
C_{i}=H_{i0}/(H_{00}-H_{ii}+EPV(i)),
\end{equation}
incorporates an infinite summation of diagrams.\cite{Ref18,Ref19} One may use a second order 
perturbation theory to evaluate the coefficients $C_{T^{+}_{k}\Phi_{i}}$ and 
$C_{R^{+}_{m}\Phi_{i}}$.\\
The vector of type 1 $T^{+}_{k}\Phi_{i}=T^{+}_{i}\Phi_{k}=T^{+}_{k}T^{+}_{i}\Phi_{0}$ can also be reached 
from $\Phi_{0}$ by other sets of excitations
$T^{+}_{m}T^{+}_{n}\Phi_{0}$ (cf Fig. 1). 
The second order expansion of the wave function tells us 
\begin{eqnarray}
C_{T^{+}_{k}\Phi_{i}} & = & \frac{H_{0k}C_{i}+H_{0i}C_{k}}{H_{00}-H_{i+k,i+k}}
\nonumber \\
& & +\frac{\sum'_{\langle m,n\rangle}
\left(
  H_{0m}C_{n}+H_{0n}C_{m}
\right)  
  }{H_{00}-H_{i+k,i+k}},
\end{eqnarray}
where $H_{i+k,i+k}=\langle\!\Phi_{T^{+}_{k}\Phi_{i}}\vert H\vert\Phi_{T^{+}_{k}\Phi_{i}}\!\rangle$ 
and $\sum'_{\langle m,n\rangle}$ runs on the couples
$T^{+}_{m}T^{+}_{n}\Phi_{0}=T^{+}_{k}T^{+}_{i}\Phi_{0}$ but different from the couple $ik$.
An infinite summation of diagrams would lead to the introduction of an $EPV$ correction in the energy 
denominator
\begin{equation}
EPV(i+k)=\underset{T^{+}_{l}T^{+}_{k}\Phi_{i}=0}{\sum_{l}}H_{0l}C_{l},
\end{equation}
and to replace $H_{00}-H_{i+k,i+k}$ by $H_{00}-H_{i+k,i+k}+EPV(i+k)$. 
\begin{equation}
C_{T^{+}_{k}\Phi_{i}} =  \frac{H_{0k}C_{i}+H_{0i}C_{k}+\sum'_{\langle m,n\rangle}(
H_{0m}C_{n}+H_{0n}C_{m})}{H_{00}-H_{i+k,i+k}+EPV(i+k)}.
\end{equation}
Now taking benefit of (Eq. 14) i.e., replacing $H_{0i}$ (and $H_{0k}$) by products 
$C_i(H_{00}-H_{ii}+EPV(i))$ in Eq. 17
one obtains an expression closer to the CC expansion:
\begin{widetext}
\begin{eqnarray}
C_{T^{+}_{k}\Phi_{i}} & = & C_{i}C_{k}
\left(
 \frac{H_{00}-H_{ii}+EPV(i)+ H_{00}-H_{kk}+EPV(k)}{H_{00}-H_{i+k,i+k}+EPV(i+k)} 
\right) 
\nonumber \\
& & +{\sum_{\langle m,n\rangle}}^{'}C_{m}C_{n}
\left(
  \frac{H_{00}-H_{mm}+EPV(m)+H_{00}-H_{nn}+EPV(n)}{H_{00}-H_{i+k,i+k}+EPV(i+k)}
\right).
\end{eqnarray}
\end{widetext}
The CC theory writes
\begin{equation}
C_{T^{+}_{k}\Phi_{i}}=C_{i}C_{k}+{\sum_{\langle m,n\rangle}}^{'}C_{m}C_{n},
\end{equation}
which ignores the possible deviation of the energy denominators from additivity i.e., assumes \\
$H_{00}-H_{i+k,i+k}=H_{00}-H_{ii}+H_{00}-H_{kk}$. 
Using the compact notation $\Delta_{i}=-H_{00}+H_{ii}$, $\Delta_{i+k}=-H_{00}+H_{i+k,i+k}$, one 
obtains 
\begin{widetext}
\begin{eqnarray}
C_{T^{+}_{k}\Phi_{i}}-C_{i}C_{k}
& = &  C_{i}C_{k}
\left(
   \frac{-\Delta_{i}+EPV(i)-\Delta_{k}+EPV(k)}{-\Delta_{i+k}+EPV(i+k)}-1 
 \right)  
 \nonumber \\
& & + \underset{T^{+}_{m}T^{+}_{n}\Phi_0=T^{+}_{k}T^{+}_{i}\Phi_0}{{\sum_{\langle m,n\rangle}}^{'}}
\hspace{-0.5cm}C_{m}C_{n}
\left(
 \frac{-\Delta_{m}+EPV(m)-\Delta_{n}+EPV(n)}{-\Delta_{i+k}+EPV(i+k)}
\right).
\end{eqnarray}
\end{widetext}
Notice that if there is only one route to create $\Phi_{\alpha}=\Phi_{i+k}$ from $\Phi_{0}$ the last 
sum disappears. If the excitation energies 
(eventually including $EPV's$) are additive, the full term vanishes in agreement with the cancellation
of disconnected diagrams. This will be the case in the 
applications to spin lattices with 
nearest-neighbor interactions when excitations 
$T^{+}_{i}$ and $T^{+}_{k}$ will concern remote bonds. \vspace{0.5cm} \\
Regarding the \textit{vectors of type2}, the operators $R^{+}_{m}(\not=T^{+})$ which lead from 
$\Phi_{i}$ to $R^{+}_{m}\Phi_{i}$  have to be treated perturbatively as follows. Analogously to 
Eq. 17 one obtains
\begin{eqnarray}
 C_{R^{+}_{m}\Phi_i} & = & 
\frac{H_{i,R_{m}^{+}i}C_{i}}{-\Delta_{R^{+}_{m}i}+EPV(R^{+}_{m}i)}
\nonumber \\
& & + \underset{R^{+}_{n}\Phi_{l}=R^{+}_{m}\Phi_{i}}{{\sum_{\langle n,l\rangle}}^{'}}
\frac{H_{l,R^{+}_{n}l}C_{l}}{-\Delta_{R^{+}_{m}i}+EPV(R^{+}_{m}i)}. 
\end{eqnarray}
\subsection{Final equations and comments}
If $R^{+}_{m}$ is a bi-electronic operator, $H_{i,R^{+}_{m}i}=H_{m}$ and the eigenequation relative 
to $\Phi_{i}$ will be
\begin{widetext}
\begin{eqnarray}
& & (\Delta_{i}-EPV(i))C_{i}+H_{i0}+\sum_{j\in S_{1}}H_{ij}C_{j} 
\nonumber \\
& & + \hspace{-0.2cm}\underset{T^{+}_{k}\Phi_{i}\not=0}{\sum_{k}}H_{0k} 
 C_{k}C_{i}
   (\frac{\Delta_{i}-EPV(i)+\Delta_{k}-EPV(k)}{\Delta_{i+k}-EPV(i+k)}-1) 
 + \hspace{-0.7cm}\underset{T^{+}_{m}T^{+}_{n}\Phi_0=T^{+}_{k}T^{+}_{i}\Phi_0}{{\sum_{\langle m,n\rangle}}^{'}}
\hspace{-0.2cm}H_{0k}C_{m}C_{n}
    (\frac{\Delta_{m}-EPV(m)+\Delta_{n}-EPV(n)}{\Delta_{i+k}-EPV(i+k)})
\nonumber \\
& & +\underset{R^{+}_{m}\Phi_{0}=0}{\sum_{m}}\frac{H_{m}^{2}C_{i}}{-\Delta_{R^{+}_{m}i}+
EPV(R^{+}_{m}i)}+\underset{R^{+}_{n}T^{+}_{l}\Phi_0=R^{+}_{m}T^{+}_{i}\Phi_0}
{{\sum_{\langle n,l\rangle}}^{'}}\hspace{-0.2cm}\frac{H_{m}H_{n}C_{l}}{-\Delta_{R^{+}_{m}i}+EPV(R^{+}_{m}i)}= 0.
\end{eqnarray}
\end{widetext}
This will be our basic equation. It is somewhat related to the usual coupled cluster equations. 
The similarity and differences appear 
clearly if ones considers the case of an SCF $\Phi_{0}$. Then the vectors of the first 
generation are all the Doubles, hence the 
method should be compared to a CCD expansion. The differences are the following:
\begin{itemize}
\item[-] the effect of the Quadruples in CCD does not take care of possible deviations from the 
additivity of the denominators and the sum over $k$ 
reduces to 
\begin{equation}
\underset{T^{+}_{k}\Phi_{i}\not=0}{\sum_{k}}C_{k}C_{i}+\underset{T^{+}_{m}T^{+}_{n}\Phi_0=
T^{+}_{k}T^{+}_{i}\Phi_0}{{\sum_{\langle m,n\rangle}}^{'}}C_{m}C_{n}.
\end{equation}
We take into account the deviation from additivity of energy denominators, 
\item[-] we take care of the coupling with the Singles and Triples through the last summation. Hence 
our method would be similar to a CCD(S,T).
\item[-] and our (semi) perturbative treatment of the Single and Triples includes $EPV$ corrections 
in the energy denominators. These corrections also appear 
in the effect of deviations from additivity of multiple excitation energies. Most of these effects 
were considered in previous works,\cite{Ref20,Ref21}
which were expressed in an intermediate Hamiltonian formalism, requiring an iterative dressing of 
small matrices. The set of coupled non linear
equations (Eq. 22) is much clearer.
\end{itemize}
\section{Applications to periodic 1-D and 2-D spin lattices}
As will be shown in this section the method is of an extreme simplicity when applied
\begin{itemize}
\item[-] to periodic lattices of spins, (and electrons, \cite{Ref22,Ref23})
\item[-] ruled by model Hamiltonians which only introduce short range interactions (usually between 
nearest-neighbor atoms)
\item[-] provided that one starts from a strongly localized wave function, product of atomic or bond 
orbitals.
\end{itemize}
The usual implementation of the Coupled Cluster method are governed by a hierarchy of excitations. 
They introduce for instance all 
Single and Double excitations. Here we adopt a different strategy, since we follow the genealogy of 
the generation of $\Psi_0$ from $\Phi_{0}$
under the successive applications of $H$. Due to the localized character of $\Phi_{0}$ and the short 
range nature of the operator in $H$, the
$1^{st}$-generation vectors will involve short range excitations only, and in periodic systems 
one has a limited number of types of 
$1^{st}$-generation excitations.
Under these conditions the number of coefficients to determine is extremely reduced, and their values 
are obtained by solving a set of 
coupled polynomial equations, which may be done on a pocket calculator. We shall take a few examples 
illustrating the method and its 
performances, comparing the so-calculated values to the exact (or highly accurate) ones. Section A 
will illustrate the derivation
of the SCP equations for simple lattices, and will compare the resulting cohesive energies to 
benchmarks values. 
Section B will employ the SCP method to study two phase transition phenomena taking place in 2-D 
lattices.
\subsection{Cohesive energy of simple spin lattices}
The ruling Hamiltonian will introduce only nearest-neighbor interactions 
\begin{equation}
H=\sum_{\langle i,j\rangle}2J(S_{i}S_{j}-1/4),
\end{equation}
between adjacent atoms. This Hamiltonian puts to zero the energy of the upper multiplet
\subsubsection{1-D chain from N\'eel}
If all bonds are equal, taking $\Phi_{0}$ as the N\'eel function (spin alternation between 
adjacent atoms), the zero order energy per atom is
$E^{(0)}_{coh}=-J$.
\begin{figure}[h]
\unitlength=1mm
\begin{picture}(80,15)
\linethickness{0.3mm}
\put(5,4){\line(1,0){60}}
\put(66,3){$\Phi_{0}$}
\put(10,5){$\uparrow$}
\put(30,5){$\uparrow$}
\put(50,5){$\uparrow$}
\put(20,5){$\downarrow$}
\put(40,5){$\downarrow$}
\put(60,5){$\downarrow$}
\put(35,1){$i$}
\end{picture}
\end{figure}
There is only one type of vectors in $S_{1}$, the one obtained from a spin exchange $T^{+}_{i}$ 
on a bond $i$, of coefficient $C$, 
coupled by $J$ with $\Phi_{0}$ ($H_{i0}=J$) hence 
\begin{eqnarray}
& & E_{coh}=-J(1-C), \nonumber \\ 
& & H_{ii}-H_{00}=\Delta_{i}=2J, \nonumber \\
& & EPV(i)=3CJ, 
\end{eqnarray}
\begin{figure}[h]
\unitlength=1mm
\begin{picture}(80,15)
\linethickness{0.3mm}
\put(5,4){\line(1,0){60}}
\put(66,3){$\Phi_{i}$}
\put(10,5){$\uparrow$}
\put(20,5){$\downarrow$}
\put(30,5){$\downarrow$}
\put(40,5){$\uparrow$}
\put(50,5){$\uparrow$}
\put(60,5){$\downarrow$}
\put(30.8,4.3){\line(1,0){10}}
\put(35,1){$i$}
\put(23,1){$h$}
\put(15,1){$g$}
\put(46,1){$j$}
\put(55,1){$k$}
\put(25,3){$/$}
\put(45,3){$/$}
\end{picture}
\end{figure}
A simple second-order estimate of the energy would lead to $C=-1/2$, $E^{(2)}_{coh}=-1.5J$, much 
larger than the exact value 
$E^{exact}_{coh}=-2J\ln2=-1.386J$. Including the $EPV's$ in the denominator would give  
\begin{eqnarray}
& & (2J-3CJ)C+J=0 \nonumber \\
& & 3C^2-2C-1=0 \nonumber \\
& &  C=-1/3 \nonumber \\
& & E'^{(2)}_{coh}=-1.333J.
\end{eqnarray}
which is already a better estimate.
There is no coupling between the vectors $\Phi_{i}$ and $\Phi_{j}\in S_{1}$. The second 
generation vectors are obtained by two spin
exchanges on bonds $i$ and $k$. If the bonds $i$ and $k$ are far the doubly excited vectors 
$T^{+}_{k}T^{+}_{i}\Phi_{0}$ have a coefficient 
$C_{i+k}=C_{i}C_{k}$ since
\begin{itemize}
\item[-] there is no other couple of $1^{st}$-generation excitation 
$T^{+}_{m}T^{+}_{n}\Phi_{0}=T^{+}_{k}T^{+}_{i}\Phi_{0}$,
\item[-] the excitation energies are additive.
\end{itemize}
The only exceptions concern the double exchanges $T^{+}_{h}T^{+}_{i}$ and $T^{+}_{i}T^{+}_{k}$, 
on second-neighbor bonds. For them 
$\Delta_{i+k}=2J$, $EVP(i+k)=5CJ$.
\begin{figure}[h]
\unitlength=1mm
\begin{picture}(80,15)
\linethickness{0.3mm}
\put(5,4){\line(1,0){60}}
\put(10,5){$\downarrow$}
\put(20,5){$\downarrow$}
\put(40,5){$\downarrow$}
\put(30,5){$\uparrow$}
\put(50,5){$\uparrow$}
\put(60,5){$\uparrow$}
\put(15,3){$/$}
\put(55,3){$/$}
\put(25,1){$i$}
\put(45,1){$k$}
\end{picture}
\end{figure}
Their coefficient can be evaluated according to (Eq. 18) as 
\begin{equation}
C_{i+l}=2C^2
\left(
 \frac{2J-3CJ}{2J-5CJ}
\right),
\end{equation}
and the final equation is 
\begin{eqnarray}
& & (2-3C)C+1+2C^2 
\left(
 \frac{4-6C-2+5C}{2-5C}
\right)=0 \nonumber \\
& & -3C^2+2C+1+2C^2
\left(
 \frac{2-C}{2-5C}
\right)=0.
\end{eqnarray}
Its solution is $C=-0.3751$ hence $E_{coh}=-1.3751J$, which deviates by $0.8\%$ from the exact value 
$-1.3862J=2J\ln2$.
\subsubsection{2-D square lattice from N\'eel}
Here $E^{(0)}_{coh}=-2J$, $E_{coh}=-2J(1+C)$, where $C$ is the coefficient relative to a spin 
exchange on a bond. Here $\Delta_{i}=H_{ii}-H_{00}=6J$, 
$EPV(i)=7CJ$. The coefficients of the doubly spin exchanged vectors are $C_{k}C_{i}$, except for
\begin{itemize}
\item[-] spin exchanges on the same plaquette.
\begin{figure}[h]
\unitlength=1mm
\begin{picture}(70,40)
\linethickness{0.3mm}
\put(5,15){\line(1,0){30}}
\put(5,25){\line(1,0){30}}
\put(15,5){\line(0,1){30}}
\put(25,5){\line(0,1){30}}
\put(15,15.3){\line(1,0){10}}
\put(15,25.3){\line(1,0){10}}
\put(10,14){$/$}
\put(30,14){$/$}
\put(10,24){$/$}
\put(30,24){$/$}
\put(14,10){$/$}
\put(14,30){$/$}
\put(24,10){$/$}
\put(24,30){$/$}
\put(20,12){$k$}
\put(20,26){$i$}
\put(11,20){$m$}
\put(26,20){$n$}
\put(37,29){since}
\put(47,29){$T^{+}_{i}T^{+}_{k}=T^{+}_{m}T^{+}_{n}$}
\put(37,19){and}
\put(45,21){$\Delta_{i+k}=8J$}
\put(45,16){$EPV(i+k)=12CJ$}
\put(33,5){hence}
\put(43,5){$C_{i+k}=2C^2
\left(
 \frac{12-14C}{8-12C}
\right)$} 
\end{picture}
\end{figure}
There are two such double excitations for a bond $i$.
\item[-] spin exchanges on other second neighbor bonds (the number of which is 14 for a given bond 
$i$).
\begin{figure}[h]
\unitlength=1mm
\begin{picture}(70,40)
\linethickness{0.3mm}
\put(5,15){\line(1,0){20}}
\put(15,5){\line(0,1){20}}
\put(25,25){\line(1,0){20}}
\put(35,15){\line(0,1){20}}
\put(25,5){\line(0,1){30}}
\put(15,15.3){\line(1,0){10}}
\put(25,25.3){\line(1,0){10}}
\put(15,25){\line(1,0){10}}
\put(25,15){\line(1,0){10}}
\put(20,24){$/$}
\put(30,14){$/$}
\put(10,14){$/$}
\put(40,24){$/$}
\put(14,10){$/$}
\put(14,20){$/$}
\put(24,10){$/$}
\put(24,30){$/$}
\put(34,20){$/$}
\put(34,30){$/$}
\put(20,12){$i$}
\put(30,22){$k$}
\put(47,31){$\Delta_{i+k}=10J$}
\put(44,16){$EPV(i+k)=13CJ$}
\put(44,5){$C_{i+k}=C^2
\left(
  \frac{12-14C}{10-13C}
\right)$}
\end{picture}
\end{figure}
\end{itemize}
The final equation is 
\begin{equation}
(6-7C)C+1+2C^2
\left(
 \frac{16-16C}{8-12C}
\right)
+14C^2
\left(
 \frac{2-C}{10-13C}
\right)=0,
\end{equation}
the solution of which is $C=-0.16327$, $E_{coh}=-2.3265J$ to be compared with the best Monte Carlo 
estimate $-2.3386J$. The error is $0.51\%$.
Our estimate may be compared with classical CC evaluations \cite{Ref24,Ref25} which give 
\begin{tabular}{cccc}
\hspace{1cm} $-2.2970J$ & with & two-body & operators \\
\hspace{1cm} $-2.3274J$ & with & four-body & operators \\
\hspace{1cm} $-2.3340J$ & with & six-body & operators \\
\hspace{1cm} $-2.3364J$ & with & eigth-body & operators. \\
\end{tabular}
\\
\noindent The use of many-body operators makes the algorithm much more complex. Our results is a 
modified two-body method, which provides the same 
accuracy as handling four-body operators. The same accuracy is obtained for anisotropic Hamiltonian. 
\subsection{Identification of phase transitions}
This section illustrates the ability of the SCPE method to locate the critical values of physical 
interactions at which phase transition occurs. 
Two examples will be considered, both concerning 2-D lattices. In all cases a relevant reference 
function $\Phi_{0}$ is selected for each 
phase, from physical arguments. The SCP equations are established, leading to an evaluation of the 
cohesive energy for the corresponding phase. 
The step-by-step derivation of these equations in not given, and the equations are reported in the 
Appendix. The intersection of the cohesive 
energy curves as functions of the structural parameter enables to give a reasonable estimate of the 
critical value of the parameters.
\subsubsection{The anisotropic Heisenberg Hamiltonian for the 2-D square lattice}
The anisotropic Hamiltonian may be written as 
\begin{equation}
H=\sum_{\langle i,j\rangle}J
\left(
 \lambda S^{i}_{z}S^{j}_{z}+S^{i}_{x}S^{j}_{x}+S^{i}_{y}S^{j}_{y}
\right). 
\end{equation}
For a 2-D square lattice and for $\lambda < -1$ the ground state is ferromagnetic, all sites bearing 
parallel spins.
For $\lambda=-1$ the ferromagnetic state is degenerate with the pure XY solution, in which the bonded
atoms bear alternatively $(\alpha+\beta)$ and
$(\alpha-\beta)$ spins. This distribution defines a reference function $\Phi_{0}$ which will be 
relevant for the $-1\leq \lambda \leq 1$ domain. 
For $\lambda \geq 1$ the wave function can be generated from the N\'eel function $\Phi'_{0}$ in which
adjacent atoms bear different spins. When $\lambda$
tends to infinity the system becomes Ising $(\Psi_0 \to \Phi'_{0})$. The phase transition between the 
XY supported and the N\'eel supported phase takes
place for the isotropic lattice ($\lambda=1$). Besides analytic series expansions \cite{Ref26,
Ref27,Ref28,Ref29} 
accurate calculations on this problem have been performed 
according to various techniques, among which one may quote
\begin{table}[t]
\caption{Cohesive energy of the anisotropic 2-D lattice.}
\begin{ruledtabular}
\begin{tabular}{cccc}
$\lambda$ &  SCP(XY) & SCP(N\'eel) & QMC \cite{Ref38}  \\
\hline
0        & -0.54672 &             & -0.54882         \\
0.1      & -0.55564 &             & -0.55681         \\
0.25     & -0.57024 &             & -0.57142         \\
0.33333  & -0.57892 &             & -0.57926         \\
0.5      & -0.59777 &             & -0.59832         \\
0.6      & -0.60965 &             & -0.60958         \\
0.75     & -0.62885 &             & -0.63017         \\
0.875    & -0.64557 &             & -0.64848         \\
0.9375   & -0.64196 &             & -0.65846         \\
0.96875  & -0.65264 &             & -0.66396         \\
1        & -0.66327 & -0.66327    & -0.66944         \\
1.02040  &          & -0.67061    & -0.67612         \\
1.11111  &          & -0.70330    & -0.70722         \\
1.2      &          & -0.73727    & -0.73920         \\
1.25     &          & -0.75696    & -0.75862         \\
1.5      &          & -0.86046    & -0.86100         \\
1.75     &          & -0.96989    & -0.96965         \\
2        &          & -1.08314    & -1.08220         \\
\end{tabular}
\end{ruledtabular}
\end{table}
\begin{figure}[b] 
\centerline{\includegraphics[scale=0.38]{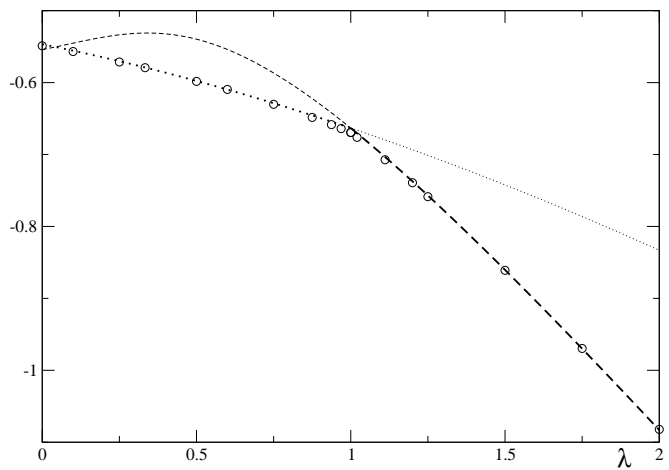}}
\caption{2-D square anisotropic spin lattice, evolution of the cohesive energy as a function of the
anisotropy parameter $\lambda$.
($\circ$) QMC, \cite{Ref38} SCP results: (...) from XY $\Phi_{0}$, (-) from N\'eel $\Phi'_{0}$.
The thick symbols identify the relevant
domains from $\Phi_{0}$ and $\Phi'_{0}$ respectively.}
\end{figure}
\begin{itemize}
\item[-] Quantum Monte Carlo calculations \cite{Ref30,Ref31,Ref32,Ref33,Ref34,Ref35,Ref36,Ref37,Ref38}
\item[-] Real Space Renormalization Group with Effective Interactions \cite{Ref39}
\item[-] Coupled Cluster expansions \cite{Ref24,Ref25}
\item[-] Dressed Cluster Method (DCM). \cite{Ref39}
\end{itemize}
The two first techniques do not introduce any reference function $\Phi_{0}$. The two last ones imply 
a choice of such reference functions either
as a zero-order function for the Coupled Cluster Method, or as a bath in which a finite cluster is 
embedded in the DCM. Our approach belongs to the 
same family. The equations derived from $\Phi_{0}=\Phi^{XY}_{0}$ or from 
$\Phi'_{0}=\Phi^{\mbox{N\'eel}}_{0}$ are given in the Appendix A. 
One may see 
that they coincide for $\lambda=1$, which appears as the critical value of the parameter. Fig. 2 
gives the evolution of the cohesive energy $E_{0}$ and
$E'_{0}$ calculated from both references and a comparison with the most accurate results. 
One sees
\begin{itemize}
\item[-]that the derivatives of the energies in $\lambda=1$ are not identical
\begin{equation}
\left(
\frac{\partial E}{\partial\lambda}\right)_{\lambda \to 1 }^{-} =
\left(
\frac{\partial E_{0}}{\partial\lambda}\right)_{\lambda = 1 }\not=
\left(
\frac{\partial E'_{0}}{\partial\lambda}\right)_{\lambda = 1 } =
\left(
\frac{\partial E}{\partial\lambda}\right)_{1 \gets \lambda }^{+},
\end{equation}
which confirms the $1^{st}$-order character of the phase transition,
\item[-] the agreement of the here-calculated cohesive energies with the best QMC calculations in 
good, the error being lower than $1\%$ in the 
$-1 < \lambda < \infty$ domain.
\end{itemize}
\subsubsection{The 1/5-depleted square lattice}
The 1/5-depleted square lattice has received interest in the context of the study of a real material, 
namely the $CaV_{2}O_{4}$ lattice. Precise 
studies of that material indicated that $2^{nd}$-neighbor interactions have to be considered. 
Nevertheless the physics of the simple lattice, built of plaquettes
and octogons, or of plaquettes connected by bonds (cf Fig. 3), appeared interesting by itself. 
There are two types of spin interactions, namely 
in the plaquette ($J_{p}$) and in the bonds between the plaquettes ($J_{d}$). Early studies did not 
give evidence of phase transitions, 
but perturbative expansions \cite{Ref40} and Quantum Monte Carlo calculations \cite{Ref41} indicate 
the existence of three phases, namely 
a gapped plaquette phase for $J_{p}/J_{d}>1.05-1.10$,
a gapped dimer phase for $J_{p}/J_{d}<0.65$ and a gapless N\'eel-type phase in the intermediate regime.
We have reexamined this problem with the SCP equations starting from three different references 
$\Phi_{0}$, namely:
\begin{enumerate}
\item a product of bond Singlets on the inter-plaquette bonds which is expected to be relevant in 
the $J_{p}/J_{d}<1$ regime
\item the N\'eel function, which should be valid in the $J_{p}\backsim J_{d}$ region
\item a product of bond Singlets on intra-plaquette bonds, relevant in the $J_{p}>J_{d}$ regime.
\end{enumerate}
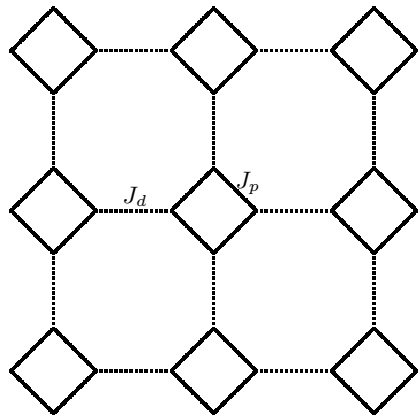
\begin{figure}[h]
\unitlength=1mm
\begin{picture}(60,55)
\linethickness{0.3mm}
\qbezier(2,26.97)(4.825,29.795)(7.65,32.62)
\qbezier(2,26.97)(4.825,24.14)(7.65,21.31)
\qbezier(7.65,32.62)(10.48,29.795)(13.31,26.97)
\qbezier(7.65,21.31)(10.48,24.14)(13.31,26.97)
\qbezier[15](7.65,32.62)(7.65,37.62)(7.65,42.62)
\qbezier[15](7.65,21.31)(7.65,16.31)(7.65,11.31)
\qbezier(7.65,42.62)(4.825,45.445)(2,48.27)
\qbezier(7.65,42.62)(10.48,45.445)(13.31,48.27)
\qbezier(2,48.17)(4.825,51.095)(7.65,53.92)
\qbezier(7.65,53.92)(10.48,51.095)(13.31,48.27)
\qbezier(7.65,11.31)(4.825,8.48)(2,5.65)
\qbezier(7.65,11.31)(10.48,8.48)(13.31,5.65)
\qbezier(2,5.65)(4.825,2.825)(7.65,0)
\qbezier(13.31,5.65)(10.48,2.825)(7.65,0)
\qbezier[15](13.31,48.27)(18.31,48.27)(23.31,48.27)
\qbezier(23.31,48.27)(26.135,51.095)(28.96,53.92)
\qbezier(23.31,48.27)(26.135,45.445)(28.96,42.62)
\qbezier(28.96,53.92)(31.785,51.095)(34.61,48.27)
\qbezier(28.96,42.62)(31.785,45.445)(34.61,48.27)
\qbezier[15](34.61,48.27)(39.61,48.27)(44.61,48.27)
\qbezier(44.61,48.27)(47.435,51.095)(50.26,53.92)
\qbezier(44.61,48.27)(47.435,45.445)(50.26,42.62)
\qbezier(50.26,53.92)(53.085,51.095)(55.91,48.27)
\qbezier(50.26,42.62)(53.085,45.445)(55.91,48.27)
\qbezier[15](13.31,5.65)(18.31,5.65)(23.31,5.65)
\qbezier(23.31,5.65)(26.135,8.48)(28.96,11.31)
\qbezier(23.31,5.65)(26.135,2.825)(28.96,0)
\qbezier(28.96,11.31)(31.785,8.48)(34.61,5.65)
\qbezier(28.96,0)(31.785,2.825)(34.61,5.65)
\qbezier[15](34.61,5.65)(39.61,5.65)(44.61,5.65)
\qbezier(44.61,5.65)(47.435,8.48)(50.26,11.31)
\qbezier(44.61,5.65)(47.435,2.825)(50.26,0)
\qbezier(50.26,11.31)(53.085,8.48)(55.91,5.65)
\qbezier(50.26,0)(53.085,2.825)(55.91,5.65)
\qbezier[15](50.26,11.31)(50.26,16.31)(50.26,21.31)
\qbezier[15](50.26,42.62)(50.26,37.62)(50.26,32.62)
\qbezier[15](28.96,11.31)(28.96,16.31)(28.96,21.31)
\qbezier[15](28.96,42.62)(28.96,37.62)(28.96,32.62)
\qbezier[15](13.31,26.97)(18.31,26.97)(23.31,26.97)
\qbezier(23.31,26.97)(26.135,29.795)(28.96,32.62)
\qbezier(23.31,26.97)(26.135,24.14)(28.96,21.31)
\qbezier(28.96,32.62)(31.785,29.795)(34.61,26.97)
\qbezier(28.96,21.31)(31.785,24.14)(34.61,26.97)
\qbezier[15](44.61,26.97)(39.61,26.97)(34.61,26.97)
\qbezier(44.61,26.97)(47.435,29.795)(50.26,32.62)
\qbezier(44.61,26.97)(47.435,24.14)(50.26,21.31)
\qbezier(50.26,32.62)(53.085,29.795)(55.91,26.97)
\qbezier(50.26,21.31)(53.085,24.14)(55.91,26.97)
\put(32,30){$J_{p}$}
\put(17,28){$J_{d}$}
\end{picture}
\caption{The 1/5-depleted 2-D square lattice.}
\end{figure}
The corresponding equations are given in Appendix B. The results appear in Fig. 4 . One should remark 
that our SCP equations never diverge, even 
when applied in paradoxical regimes, for instance when one enters in the $J_{p} \ll J_{d}$ regime 
from inter-plaquette bonds. 
\begin{figure}[h]
\centerline{\includegraphics[scale=0.38]{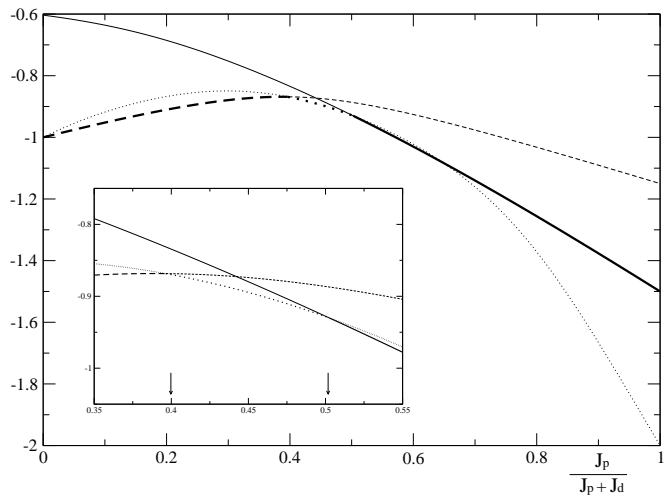}}
\caption{1/5-depleted 2-D square lattice. Evolution of the cohesive energy as a function of the
$J_{p}/(J_{p}+J_{d})$ ratio according to the 
SCP method. (- -) from inter-plaquette bond Singlets $\Phi_{0}$, (...) from N\'eel $\Phi'_{0}$, and
(---) from plaquette Singlets  $\Phi''_{0}$.
The thick symbols identify the relevant domains from the various reference functions. The arrows
indicate the phase transition critical values.}
\end{figure}
The expansion from N\'eel tends to an erroneous ($-2J$ instead of $-1.5J$) value for $J_d=0$ (isolated plaquettes). Using the 
perturbative evaluation of the coefficient of the doubly permuted vectors (Eq. 17 instead of 
Eq. 18), one reaches the correct asymptotic value for $J_d=0$. One must concentrate
on the estimates of the cohesive energy in the relevant domains. As seen from Fig. 4 one notices 
the existence of three phases.
\begin{itemize}
\item[-] a dimer phase for $J_{p}/J_{d}<0.6563$
\item[-] a N\'eel-order phase for $0.65<J_{p}/J_{d}<1.0105$
\item[-] a plaquette phase for $J_{p}/J_{d}>1.0105$.
\end{itemize}
Using Eq. 17 instead of Eq. 18, the critical ratios remain almost identical.
The method confirms the existence of these three phases and the calculated critical values of the 
$J_{p}/J_{d}$ ratios are in good agreement with 
previously mentioned estimates.
\section{Discussion and Conclusion}
The here presented method is based on perturbative expansion of the wave-function but it is not 
strictly perturbative. It is 
perturbative in the sense that it follows a genealogic generation of the wave function and that it 
uses (or refers to) pertubative estimates of the 
coefficients of the second generation vectors.\\
But it is not perturbative in the sense that the coefficients of the first generation vectors 
are obtained by solving a set of coupled polynomial equations. It may compared to the  
Coupled Cluster method (and in particular to a CCD(S,T) version) from which it differs by the 
consideration of possible deviations 
from additivity of multiple-excitation energies and by the introduction of $EPV$ corrections in 
energy denominators.\\
Although perfectly applicable to \textit{ab initio} calculations, the method has been essentially designed 
for the study of periodic
lattices and model Hamiltonians. In such cases, starting from strongly localized zero-order pictures, 
the method is analytical, it
leads to simple polynomial expressions, introducing a very limited number of variables (one per type 
of bond in most cases), which 
can be solved on a pocket calculator. The accuracy of such a simple method is surprising, as 
illustrated above. It provides an 
elegant exploratory tool for the study of spin periodic lattices. The present work has shown that 
the method may be employed to 
identify phase transitions. The method is of course applicable 
to Hubbard Hamiltonians, 
whose exact solutions are not known (except for the 1-D regular chain) and to 2-D or 3-D lattices. 
Since the method only explores 
the physics around a given bond (up to six bonds in each direction) it may lead to some stimulating 
qualitative discussions 
to assess what is local, regional and properly collective in the cohesive energy of delocalized 
systems. \cite{Ref22,Ref23} Further work will study the spin-Peierls distortions.
The present method essentially furnishes an estimate of the ground state energy. It may also provide
short range correlation functions. The localized character of $\Phi_0$ and the fact that the 
excitation processes playing a role in the low-order processes only concern a few bonds prevent to 
give estimates of the correlation beyond this limit. The related Coupled Cluster formalism has led 
to specific methods for the evaluation of spectra, namely the Equation of Motion Coupled Cluster
(EOM-CC) algorithm. \cite{Ref42} Future works will analyse the possibility of such an extension in the 
frame of our development.
\begin{acknowledgments}
The authors thank D. Poilblanc and S. Capponi for helpful discussions.
\end{acknowledgments}
\begin{widetext}
\appendix
\section{}
SCP equations for the anisotropic 2-D square lattice (Eq. 30)
\begin{enumerate}
\item equation from N\'eel
\begin{equation}
\left(
 6\lambda J-7JC\right)C+J+2J
\left[
2
\left(
  \frac{12\lambda J-14JC}{8\lambda J-12JC}\right)-1\right]C^2
+14J
\left[
\frac{12\lambda J-14JC}{10\lambda J-13JC}-1\right]C^2 = 0
\end{equation}
\begin{equation}
E_{coh}=J(C-\frac{\lambda}{2})
\end{equation}
\item equation from XY
\begin{eqnarray}
&&
\left[
 6J -\frac{7J}{2}
 \left( 
 1+\lambda\right)C\right]C+\frac{J}{2}
 \left(
 1+\lambda\right)+2J
 \left[
 2
 \left(
 \frac{12J-7J(1+\lambda)C}{8J-6J(1+\lambda)C}\right)-1\right]C^2 
 \\ \nonumber
 & & +14J
\left[
 \frac{12J-7J(1+\lambda)C}{10J-\frac{13J}{2}(1+\lambda)C}-1\right]C^2
 -\frac{J^2(1-\lambda)^2C}{2
 \left[
 J-\frac{J}{2}(1+\lambda)C\right]}-\frac{J^2(1-\lambda)^2C}{8
 \left[
 J-\frac{J}{2}(1+\lambda)C\right]}=0
\end{eqnarray}
\begin{equation}
E_{coh}=\frac{J}{2}
\left[
(1+\lambda)C-1\right]
\end{equation}
\end{enumerate}
\section{}
SCP equations for the 1/5-depleted 2-D square lattice with $J_{p}=(1+\delta)$ and $J_{d}=(1-\delta)$
\begin{enumerate}
\item equation from inter-plaquette bond Singlets
\begin{equation}
\left[
4(1-\delta)-\frac{5}{6}(1+\delta)-\frac{7\sqrt{3}}{2}(1+\delta)C\right]C+\frac{\sqrt{3}}{2}(1+\delta)
-\frac{15(1+\delta)^2C}{4(1-\delta)-\frac{9\sqrt{3}}{2}(1+\delta)C}=0
\end{equation}
\begin{equation}
E_{coh}=\frac{1}{2}
\left[
-\frac{3}{2}+\frac{\delta}{2}+\frac{\sqrt{3}}{2}(1+\delta)C\right]
\end{equation}
\item equations from N\'eel function. The coefficient $C$ (resp. $C'$) is relative to the spin 
exchange in the plaquette (resp. between the plaquettes)  
\begin{eqnarray}
&&
\left[
4-3(1+\delta)C-2(1-\delta)C'\right]C+(1+\delta)+(1+\delta)
\left[
2\frac{8-6(1+\delta)C-4(1-\delta)C'}{4(1-\delta)-4(1-\delta)C'-4(1+\delta)C}-1\right]C^2 
\\ \nonumber
&&
4(1-\delta)
\left[
\frac{8-6(1+\delta)C-4(1-\delta)C'}{4(1+\delta)+2(1-\delta)-4(1+\delta)C-3(1-\delta)C'}-1\right]C^2
\\ \nonumber
&& 
+2(1+\delta)
\left[
\frac{4+4(1+\delta)-7(1+\delta)C-3(1-\delta)C'}{4(1+\delta)+2(1-\delta)-4(1+\delta)C-3(1-\delta)C'}-1\right]CC'
=0
\end{eqnarray}

\begin{eqnarray}
&&
\left[
4(1+\delta)-4(1+\delta)C-(1-\delta)C'\right]C'+(1-\delta)
\\ \nonumber
&&
+4(1+\delta)
\left[
\frac{4+4(1+\delta)-7(1+\delta)C-3(1-\delta)C'}{4(1+\delta)+2(1-\delta)-4(1+\delta)C-3(1-\delta)C'}-1\right]CC'
\\ \nonumber
&&
4(1+\delta)
\left[
\frac{8(1+\delta)-8(1+\delta)C-2(1-\delta)C'}{6(1+\delta)-6(1+\delta)C-2(1-\delta)C'}-1\right]C'^2=0
\end{eqnarray}
\begin{equation}
E_{coh}=\frac{1}{2}
\left[
-\frac{3}{2}-\frac{\delta}{2}+(1+\delta)C+\frac{1}{2}(1-\delta)C'\right]
\end{equation}
\item equations from intra-plaquette bond Singlets. $C$ and $C'$ concern respectively intra and 
inter-plaquette diexcitations
\begin{equation}
\left[
2(1+\delta)-2\sqrt{3}(1-\delta)C'+(1+\delta)C\right]C+\sqrt{3}(1+\delta)
+ \frac{2\sqrt{3}(1-\delta)
\left[
\frac{\sqrt{3}}{2}(1-\delta)C+\sqrt{3}(1+\delta)C'\right]}{-4(1+\delta)+2\sqrt{3}
\left[
(1-\delta)C'+(1+\delta)C\right]}=0
\end{equation}
\begin{eqnarray}
&&
\left[
\frac{19}{6}(1+\delta)-\frac{3\sqrt{3}}{2}(1-\delta)C'-2\sqrt{3}(1+\delta)C\right]C'+\frac{\sqrt{3}}{2}(1-\delta)
\\ \nonumber
&&
+\frac{3(1-\delta)^2C'+2\sqrt{3}(1+\delta)
\left[
\frac{\sqrt{3}}{2}(1-\delta)C+\sqrt{3}(1+\delta)C'\right]}{-4(1+\delta)+2\sqrt{3}
\left[
(1-\delta)C'+(1+\delta)C\right]}=0
\end{eqnarray}
\begin{equation}
E_{coh}=\frac{1}{2}
\left[
-\frac{3}{2}-\delta+\frac{\sqrt{3}}{4}
\left[
(1+\delta)C+(1-\delta)C'\right]\right]
\end{equation}
\end{enumerate}
\end{widetext}

\end{document}